Time-Series Analysis of Super-Kamiokande Measurements of the Solar Neutrino Flux

P.A. Sturrock

Center for Space Science and Astrophysics, Varian 302G,
Stanford University, Stanford, CA 94305

ABSTRACT

The Super-Kamiokande Consortium has recently released data suitable for time-series analysis. The binning is highly regular: the power spectrum of the acquisition times has a huge peak (power S > 120) at frequency 35.98 $y^{-1}$ (period 10.15 days), where power measurements are such that the probability of obtaining a peak of strength S or more by chance at a specified frequency is exp(-S). This inevitably leads to severe aliasing of the power spectrum. The strongest peak in the range 0 – 100 in a power spectrum formed by a likelihood procedure is at 26.57 $y^{-1}$ (period 13.75 days) with S = 11.26. For the range 0 – 40 $y^{-1}$, the second-strongest peak is at 9.42 $y^{-1}$ (period 38.82 days) with S = 7.3. Since 26.57 + 9.42 = 35.99, we conclude that the weaker peak at 9.42 $y^{-1}$ is an alias of the stronger peak at 26.57 $y^{-1}$. We note that 26.57 $y^{-1}$ falls in the band 26.36 – 27.66 $y^{-1}$, formed from twice the range of synodic rotation frequencies of an equatorial section of the Sun for normalized radius larger than 0.1. Oscillations at twice the rotation frequency, attributable to "m = 2" structures, are not uncommon in solar data. We find from the shuffle test that the probability of obtaining a peak of S = 11.26 or more by chance in this band is 0.1 %. This new result therefore supports at the 99.9% confidence level previous evidence, found in Homestake and GALLEX-GNO data, for rotational modulation of the solar neutrino flux. The frequency 25.57 $y^{-1}$ points to a source of modulation at or near the tachocline.

1. INTRODUCTION

In recent articles, we have presented the results of our analyses of Homestake (Davis & Cox 1991; Lande et al. 1992; Cleveland et al. 1995, 1998) and GALLEX-GNO (Anselmann et al. 1993, 1995; Hampel et al. 1996, 1997; Altman et al. 2000) solar neutrino data. The main focus in these articles has been the search for evidence of modulation that may be attributed to solar rotation. In GALLEX-GNO data, we find a modulation at 13.59 $y^{-1}$ (period 26.88 days), corresponding to a sidereal rotation rate of 14.59 $y^{-1}$ (462 nHz), which matches the rotation rate as determined by helioseismology (Schou et al. 1998) at or near the equator in the convection zone at a normalized radius of about 0.85 (Sturrock & Weber 2002). In Homestake data, we find the same modulation but we also find a stronger modulation at 12.88 $y^{-1}$ (period 29.1 days), corresponding to a sidereal rotation frequency of 13.88 $y^{-1}$ (440 nHz), compatible with estimates of the rotation rate of the radiative zone made by helioseismology (Sturrock, Walther, & Wheatland 1997).

Neutrino experiments that use Cerenkov detectors, namely Kamiokande (Suzuki 1995; Fukuda et al. 1996), Super-Kamiokande (Fukuda et al. 2001, 2002) and SNO (Ahmad et al. 2002a, 2002b), have a much higher event rate, and events are timed very precisely, so they have the potential of providing a definitive answer to the question of variability on the time-scale of solar rotation. Until recently, this question could not be answered since data sets published by these consortia has typically comprised a few data points, each representing an average over several weeks or longer. However, the Super-Kamiokande consortium has recently released data organized in bins of only a few days duration (Smy 2002). Early analyses of this data set by Sturrock and Caldwell (2003) and by Milsztajn (2003) indicate that this data set yields evidence of rotational modulation.

The timing of the Super-Kamiokande data is highly regular, so we must expect aliasing problems. For this reason, before analyzing the flux data, we examine in Section 2 the timing by forming a power spectrum of the midpoints of the bins. We find a huge peak at the frequency $\nu = 35.98$ (period 10.15 days), where frequencies are measured in cycles per year. We derive a power spectrum of the flux measurements in Section 3, using a likelihood procedure similar to that used in our analysis of Homestake data (Sturrock, Walther, & Wheatland, 1997). The strongest feature in this spectrum is a peak a $\nu = 26.57$, but we find other peaks that appear to be aliases of that frequency. We therefore address the aliasing issue in Section 4. We present a significance estimate in Section 5, a map indicating the probable source of the modulation in Section 6, and further discussion in Section 7.

## 2. TIMING ANALYSIS

The published Super-Kamiokande data comprises averages over 184 bins, each about 10 days in duration, from the time frame May 1996 to July 2001. We enumerate the bins by the index r, r = 1, 2, …, R, where R = 184. For each bin, we are given the start time $t_{sr}$ and the end time $t_{er}$. We have, for convenience, converted the times to years. We are also given the estimated flux (in SNU) and the normalized Sun-Earth distance, from which we may compute the flux as it would be measured at 1AU, that we denote by $g_r$. Since the probability distribution function for the flux is somewhat asymmetric, we are given both upper and lower 1-sigma error estimates, that we denote by $\sigma_{ur}$ and $\sigma_{lr}$, respectively.

In order to anticipate what aliasing may occur, we need to examine the power spectrum of the timing. For this purpose, we form the mean time $t_{mr}$ of each run:

$$t_{mr} = \tfrac{1}{2}(t_{sr} + t_{er}), \qquad (2.1)$$

and consider a time series that has delta functions at these times:

$$x(t) = \sum_{r=1}^{R} \delta(t - t_{mr}). \qquad (2.2)$$

Then the Rayleigh power (see, for instance, Mardia 1972) of this time series is given by

$$S_T(\nu) = \frac{1}{R}\left|\sum_{r=1}^{R} e^{i2\pi\nu t_{mr}}\right|^2. \qquad (2.3)$$

This spectrum is shown in Figure 1. We see that there is a huge peak with power $S = 123.5$ at $\nu_T = 35.98$, corresponding to a period of 10.15 days. The next highest peak has power $S = 50.5$ at $\nu = 72.00$, which is effectively the harmonic of $\nu_T$. There is also a peak (not shown) with power $S = 40$ at $\nu = 3\nu_T$, Hence if there is a real peak in the spectrum at frequency $\nu_0$, we may expect to find peaks also at the primary alias frequencies $\nu_0 + \nu_T$ and $|\nu_0 - \nu_T|$, smaller peaks at the secondary alias frequencies $\nu_0 + 2\nu_T$ and $|\nu_0 - 2\nu_T|$, etc.

## 3. POWER SPECTRUM ANALYSIS

We now compute the power spectrum of the data by a technique similar to that previously used in our analysis of Homestake data (Sturrock, Walther, & Wheatland, 1997). In that calculation, we examined fits of the data to a three-parameter functional form for the neutrino flux:

$$f(t) = C + Ae^{i2\pi\nu t} + A^* e^{-i2\pi\nu t} . \tag{3.1}$$

However, we have found from a re-examination of our early analysis of GALLEX data (Sturrock et al. 1999) that there is a risk that power spectra formed in this way may be contaminated by any periodicity in the timing data. Since, as we have just seen, the timing of Super-Kamiokande data has an exceedingly strong periodicity, we prefer to avoid the above procedure.

For this reason, we consider only fluctuations of the flux relative to the mean flux $g_m$, and therefore consider the variable

$$x_r = \frac{g_r}{g_m} - 1 . \tag{3.2}$$

The Super-Kamiokande data set lists upper and lower error estimates for each run but, for our purposes, the difference between the upper and lower estimates is not significant, and we therefore introduce a single error estimate corresponding to the variable $x_r$ as follows:

$$\sigma_r = \frac{\sigma_{ur} + \sigma_{lr}}{2g_m}. \tag{3.3}$$

If f(t) represents the neutrino flux (multiplied by a capture coefficient) as a function of time, then the measurements are given by

$$g_r = \frac{1}{D_r} \int_{t_{sr}}^{t_{er}} dt\, f(t), \tag{3.4}$$

where $D_r$ is the duration of each run:

$$D_r = t_{er} - t_{sr}. \tag{3.5}$$

If, by analogy with (3.2), we write

$$f(t) = g_m (1 + \phi(t)), \tag{3.6}$$

we find that the expected values of $x_r$ are given by

$$X_r = \frac{1}{D_r} \int_{t_{sr}}^{t_{er}} dt\, \phi(t). \tag{3.7}$$

We now consider the following functional form for $\phi(t)$,

$$\phi(t) = A e^{i 2\pi \nu t} + A^* e^{-i 2\pi \nu t}, \tag{3.8}$$

so that |A| will be one half the depth of modulation of the neutrino flux. Then the log-likelihood that the data may be fit to the above functional form for $\phi(t)$ is given (apart from an additive constant term) by

$$L = -\tfrac{1}{2}\sum_{r=1}^{R}\frac{(x_r - X_r)^2}{\sigma_r^2}, \qquad (3.9)$$

where

$$X_r = \frac{1}{D_r}\int_{t_{sr}}^{t_{er}} dt\left(Ae^{i2\pi vt} + A^* e^{-i2\pi vt}\right), \qquad (3.10)$$

and, for each frequency, A adjusted to maximize the likelihood (Sturrock, Walther, & Wheatland 1997).

What is significant is not the actual likelihood as computed from (3.9), but the increase in the log-likelihood over the value that corresponds to a constant flux, which is given (apart from the same constant term) by

$$L_0 = -\tfrac{1}{2}\sum_{r=1}^{R}\frac{x_r^2}{\sigma_r^2}. \qquad (3.11)$$

Hence the relative log-likelihood is given by

$$S = L - L_0, \qquad (3.12)$$

i.e. by

$$S = \tfrac{1}{2}\sum_{r=1}^{R}\frac{x_r^2}{\sigma_r^2} - \tfrac{1}{2}\sum_{r=1}^{R}\frac{(x_r - X_r)^2}{\sigma_r^2}. \qquad (3.13)$$

We use the symbol S since the quantity we have computed is distributed in the same way as the power computed for a normally distributed random variable, for which the probability of obtaining a value in the range S to S+ dS by chance is $e^{-S}dS$ or, equivalently, the probability of obtaining the value S or larger is $e^{-S}$.

We show in Figure 2 the power spectrum over the frequency range 0 to 40 $y^{-1}$, corresponding to periods of about 9 days and longer. This covers the range of interest in a search for the effects of solar rotation, for which one might expect an oscillation with period of order 27 days or (as often happens in solar research) an oscillation with period of order 13.5 days. We see that the highest peak in that range is at the frequency 26.57 y$^{-1}$, corresponding to a period of 13.75 days, for which S = 11.26. The probability of obtaining a peak this large or larger by chance at a specified frequency is about one part in 10$^5$. However, the solar rotation frequency does not have a unique value: it varies strongly with latitude and radius, and may also varies with time. Hence we need to examine the probability of obtaining by chance such a feature that may be related to the Sun's internal rotation. Before we take up this question, however, we need to be sure that the peak at 26.57 y$^{-1}$ is the primary periodicity in the time series. We therefore take up the question of aliasing in Section 4, and postpone the significance calculation until Section 5.

## 4. ALIASING

In order to examine the role of aliasing in our power spectrum analysis, it is convenient to examine the spectrum over a wider frequency range. We show in Figure 3 the power spectrum over the frequency range 0 to 100 y$^{-1}$. We see that, over this range, the strongest peak is that at frequency 26.57 y$^{-1}$ with power 11.3. The next three peaks are, in order, (a) $\nu = 62.56$ with S = 10.02, (b) $\nu = 84.75$ with S = 7.61, and (c) $\nu = 9.42$ with S = 7.29.

If we denote by $\nu_0$ the frequency of the strongest peak at 26.57 y$^{-1}$, and if we recall from Section 2 that the timing spectrum has its major peak at $\nu_T = 35.98$, we see that we may identify (a) as an expected peak at $\nu_T + \nu_0$ and (c) as an expected peak at $|\nu_T - \nu_0|$. We also find a peak at 98.56 with S = 6.2, that we may identify as an expected peak at $2\nu_T + \nu_0$, and a peak with power 3.8 at frequency $\nu = 45.42$ that we may identify with an expected peak at $|2\nu_T - \nu_0|$. Hence if we identify $\nu_0$ as the fundamental periodicity in the time series, we may recognize in the spectrum several of the expected alias frequencies.

We have examined the possibility that the primary periodicity may be one of those listed as (a) to (c) in the preceding paragraph, but this seems less likely. For instance, if we consider the possibility that the fundamental periodicity is that at 9.42 y$^{-1}$, for which S = 7.3, we would find that the first-order alias frequencies have powers 11.3 (at $\nu = 26.57$) and 3.6 (at $\nu = 45.40$). Since we expect the powers of the two first-order aliases to be comparable, and both to be less than the power of the fundamental periodicity, this option is unsatisfactory. We therefore explore further the possibility that the fundamental periodicity is that at $\nu = 26.57$.

We have determined the amplitude and phase of the sine wave with frequency 26.57 y$^{-1}$ that provides the best fit to the data. We find that $|A| = 0.05$, corresponding to a 10% depth of modulation. We have subtracted this sine wave from the data and recomputed the power spectrum, as shown in Figure 4. Whereas for the full range 0 to 100 y$^{-1}$ we see seven peaks with S > 6 in Figure 3, we see only two in Figure 4. For the smaller range 0 to 40 y$^{-1}$, we see six peaks with S > 5 in Figure 3, but we see none in Figure 4. Hence the "CLEAN" spectrum (see, for instance, Roberts, Lehar & Dreher 1987), formed by removing the influence of the fundamental oscillation, is indeed "cleaner" than the original spectrum.

As a further guide to the influence of the fundamental oscillation on the power spectrum, we show in Figure 5 the difference between the original spectrum shown in Figure 3 and the CLEAN spectrum shown in Figure 4. This clearly shows that the CLEAN process has removed not only the fundamental, but also the alias frequencies at $|\nu_T - \nu_0|$, $\nu_T + \nu_0$, $|2\nu_T - \nu_0|$, $2\nu_T + \nu_0$, and $|3\nu_T - \nu_0|$. (A peak at $3\nu_T + \nu_0$ would be outside the range 0 to 100 y$^{-1}$.)

## 5. SIGNIFICANCE TEST

It is important to estimate the significance of the peak at frequency 26.57 y$^{-1}$ in the power spectrum. If the frequency were specified in advance, the probability of obtaining a peak this large (S = 11.26) or larger would be given by e$^{-S}$, i.e. 1.3 10$^{-5}$. However, the frequency was not specified in advance, so this estimate is inappropriate. Since our purpose is to test the hypothesis that the feature at 26.57 $y^{-1}$ is due to solar rotation, the appropriate search band is determined by the range of rotation frequencies in the solar interior.

According to Bahcall (1989), the $^8$B neutrinos detected by Cerenkov experiments are produced almost entirely within a sphere of normalized radius 0.1. Taking into account this fact and the fact that the axis of rotation of the Sun has an inclination of only about 7º to the normal to the ecliptic, we see that most of the neutrinos detected on Earth travel close to the equatorial section of the Sun. It is therefore adequate for our purposes to consider the solar rotation rate in that section. The internal rotation of the Sun has been estimated from helioseismology. According to the analysis presented by Schou et al. (1998), the sidereal rotation frequency in an equatorial section of the Sun, for $0.1 \leq r \leq 1$, extends over the range 13.68 to 14.83 $y^{-1}$. Hence the range of the equatorial synodic rotation frequencies is 12.68 to 13.83 $y^{-1}$. It is not unusual for solar features and events to show a stronger periodicity at the harmonic of the rotation frequency than at the rotation frequency itself (see, for instance, Donnelly et al. 1983; Pap, Tobiska, & Bouwer, 1990; Mursula & Zieger, 1996), due to the tendency of active regions to form in pairs with 180 degree separation (Bai 1987, 1989, 2002). The appropriate range for the harmonic is 25.36 to 27.66 $y^{-1}$, and we have adopted this as the "search band" for our significance estimate.

The "shuffle test" (see, for instance, Bahcall & Press 1991) provides a robust significance estimate. The procedure is to retain the start-time and end-time pairs as one part of the data, and the flux measurements and the upper and lower error estimates as the other part of the data, and then randomly reassign the second part to the first part. If the measurements were due to a stationary random process, shuffling should have no effect on the spectrum.

We show in Figure 6 the distribution of values of the maximum power in the search band from 10,000 simulations formed by the shuffle process outlined above. We see that only 8 have a power of 11.26 or more in the search band, from which we may conclude that the peak at 26.57 y$^{-1}$ is significant, as an indicator of internal rotation, at the 99.9% confidence level.

## 6. ROTATIONAL MODULATION STATISTIC

Following the procedure introduced in our analysis of GALLEX-GNO data (Sturrock & Weber 2002), we now develop a visual display of the relationship between the periodicity of the solar neutrino flux, as measured by the Super-Kamiokande experiment, and the Sun's internal rotation. Schou et al. (1998) have tabulated the rotation rate $v_h(r,\lambda)$ and the error estimate $\sigma_h(r,\lambda)$ for 101 values of the radius $r$ and 25 values of the latitude $\lambda$, deriving their estimates from the MDI helioseismology experiment on the SOHO spacecraft. These data determine a probability distribution function of the rotation frequency for each pair of values $r, \lambda$:

$$P(v|r,\lambda) = (2\pi)^{-1/2} \sigma_h^{-1} \exp\left[-\frac{(v-v_h)^2}{2\sigma_h^2}\right]. \tag{6.1}$$

Since we wish to compare the rotation frequencies with the neutrino-flux variability as measured on Earth, it is appropriate to inspect the synodic rotation rates rather than the sidereal rates.

We also need to take account of the fact that the Super-Kamiokande measurements appear to respond to an m = 2 structure within the Sun. We therefore modify our earlier definition of the rotational modulation statistic as follows

$$\Xi(r,\lambda) = \int_{V_a}^{V_b} dv\, S(2v) P(v|r,\lambda), \tag{6.2}$$

where S is the spectrum computed in Section 3 and displayed in Figures 2 and 3. In this integral, we need to select limits of integration that are sufficiently wide to cover all significant contributions from the PDF. We have adopted $v_a = 0$ and $v_b = 20$, but a much smaller range would have been satisfactory. A color map of $\Xi$ as a function of radius and latitude is presented in Figure 7.

This figure is essentially a mapping of the power spectrum of the solar-neutrino time series onto the solar interior. Where the map is colored yellow or red, $\Xi$ is large compared with its average value, denoting a "resonance" between the neutrino flux and the harmonic of the local solar rotation. Another way of looking at this figure, which is perhaps physically more significant, is the following: Let us assume that there is a well-defined oscillation in the neutrino flux, and let us assume that this oscillation is due to modulation of the flux by an m = 2 structure (such as a double magnetic structure) in the solar interior. Then we can attempt to locate that structure by finding the location (or locations) where the rotation rate has just the correct value to account for the dominant oscillation of the neutrino flux. The map shown in Figure 7 may now be viewed as a probability distribution function for the location of the modulating structure. Since neutrinos detected on Earth have traveled fairly close to an equatorial section of the Sun, we see that the modulation appears to take place at a normalized radius of about 0.7, i.e. at or near the tachocline.

However, the above interpretation is predicated on the assumption that the neutrino modulation is due to magnetic structures that have the same rotation rate as is determined by helioseismology, and we should point out caveats to this assumption. It is possible that the internal rotation is variable (due perhaps to a localized and transient "jet stream") on length-scales and time-scales that do not easily show up in helioseismology data. It is also possible that the Sun's internal magnetic field exhibits MHD oscillations. For instance, an Alfven wave traveling around a toroidal flux tube could lead to modulation at the rotation frequency of the traveling wave. As a result, modulation of the solar neutrino flux with a frequency of about $26.6 y^{-1}$ could be caused by a retrograde Alfven wave with an $m = 2$ structure propagating in a flux tube of field strength 2,000 gauss at normalized radius 0.85. The simplest interpretation of modulation of the solar neutrino flux may not be the correct interpretation.

# 7. DISCUSSION

The main result of this analysis is that Super-Kamiokande neutrino flux measurements exhibit an oscillation with frequency 26.57 y$^{-1}$ corresponding to a period of 13.75 days. This is not an unusual result in solar physics. For instance, Pap, Tobiska, and Bouwer (1990) have found periods in the range 13.7 to 13.8 days in 42 years of solar 10.7 cm radio flux measurements, 17-years of Ca-K measurements, 8 years of projected areas of active spots, and 108 years of the sunspot blocking function. Since solar activity has its origin in the solar magnetic field, these analyses indicate that part of the Sun's internal magnetic field has an m = 2 structure that rotates with a synodic frequency of about 13.3 y$^{-1}$, corresponding to a sidereal frequency of 14.3 y$^{-1}$ or about 450 nHz, which (as we noted in Section 6) points to a source of modulation of the solar neutrino flux at or near the tachocline.

In order to examine the empirical "wave-form" of the flux measurements, we show in Figure 7 the values g$_r$ as a function of phase defined by

$$\phi_r = \mod(t_{m,r} - 1970, 1) \ . \tag{7.1}$$

The sine-wave modulation is quite evident in this figure, which also shows the best-fit sine wave. The phase agrees with that of the best fit to the actual flux, as found from the spectrum analysis of Section 3. However, the depth of modulation of the flux measurements is only 3%, whereas (as we noted in Section 4) the depth of modulation of the flux is 10%. The difference is due to the fact that measurements are formed by integrating the flux over bins of duration 10 days, approximately. As a result, the depth of modulation is reduced by the factor

$$\frac{DOM(g)}{DOM(f)} = \frac{\left|1 - e^{i2\pi\nu D}\right|}{2\pi\nu D} \ . \tag{7.2}$$

For $\nu = 26.57\,y^{-1}$ and $D = 0.027\,y$, the reduction factor is 0.32, leading to an expected depth of modulation of 3% for the flux measurements, in agreement with the depth of modulation shown in Figure 7.

It is clear that the spectrum analysis of Section 3 would have yielded a clearer result if the data had not been binned, or if the bin widths had been shorter than 10 days. If data from Cerenkov experiments could in future be packaged into one-day bins, preferably tied to Universal Time (UT), the essential characteristics of the variation would be preserved, and it would be possible for any interested particle physicist or solar physicist to carry out time-series analysis of the data. Standardization would facilitate the comparison of data from different Cerenkov experiments, and this particular choice would have the merit that the data could be compared with other solar data sets, which are frequently packaged in this way.

The data could conveniently be expressed as a probability distribution function for the number of captures in each time bin, or as a probability distribution function for the mean flux for each bin. The results of radiochemical solar neutrino experiments may similarly be expressed as probability distribution functions either for the number of captures or for the survival-weighted flux in each run. Such formatting of radiochemical and Cerenkov data would also facilitate the "global" analysis of all neutrino data sets.

Thanks are due to the Super-Kamiokande consortium for making their data available, and to Taeil Bai, David Caldwell, Alexander Kosovichev, Jeffrey Scargle, and Guenther Walther for their interest and for helpful discussions. This work was supported by NSF grant AST-0097128.

FIGURES

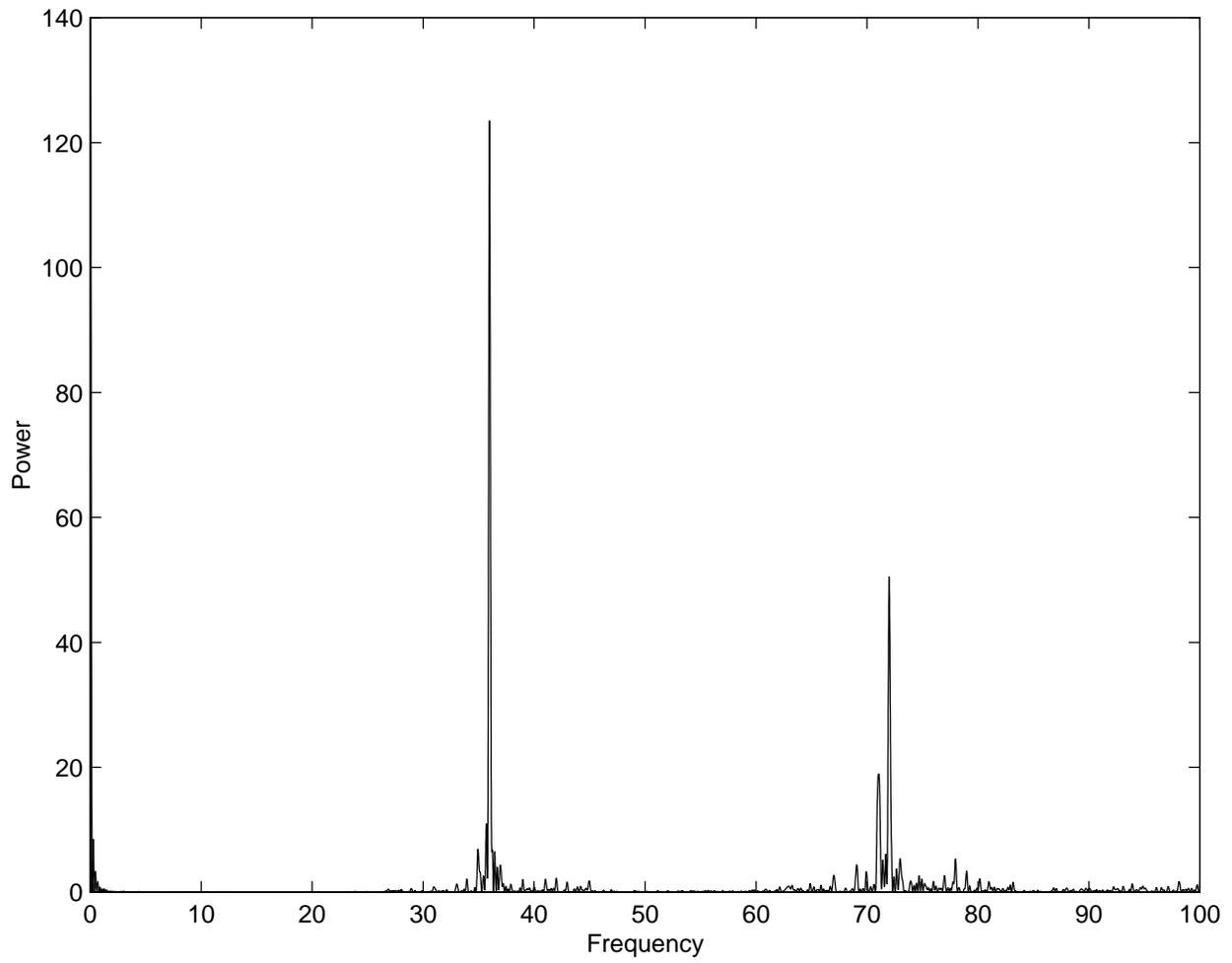

Figure 1. The power spectrum formed from the mean times of the data bins.

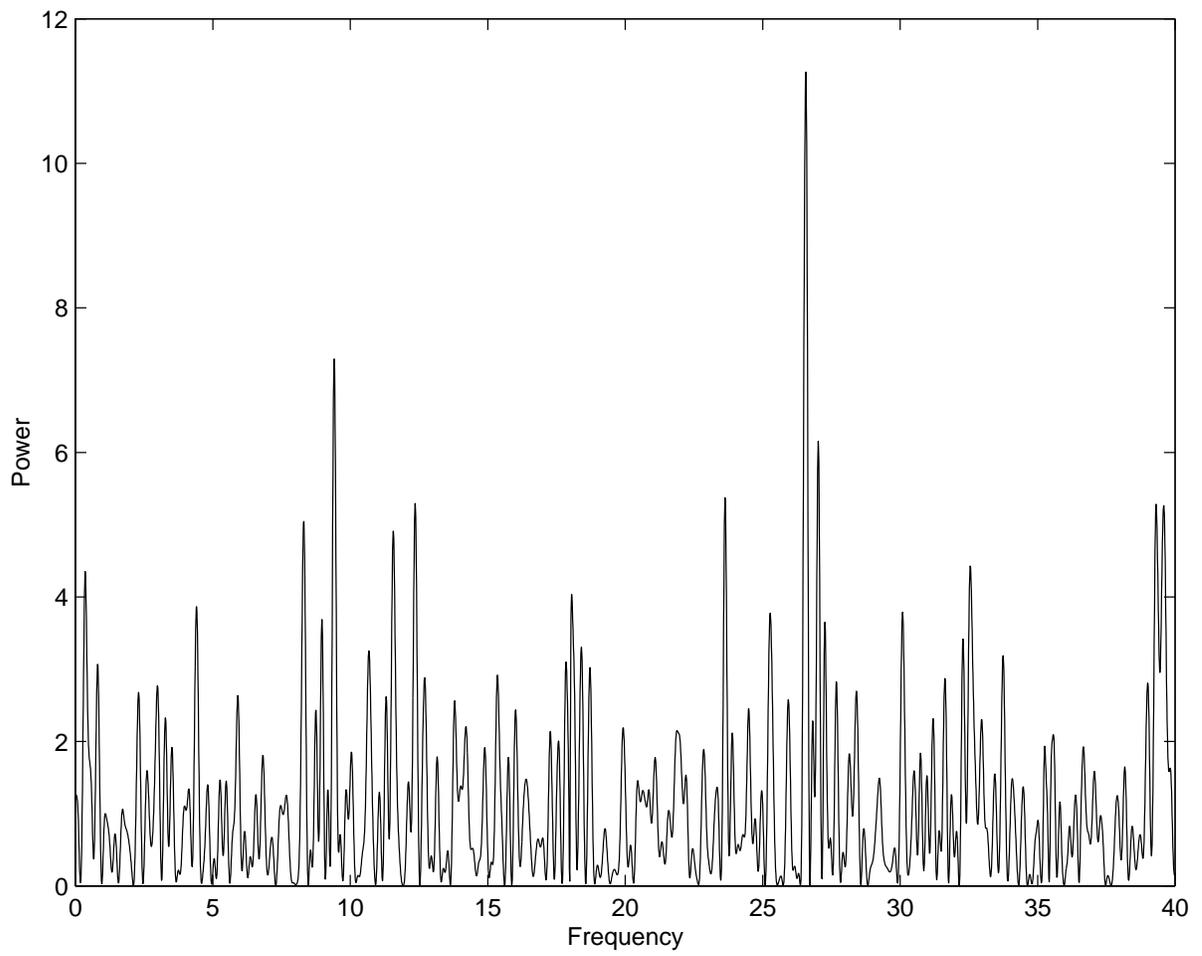

Figure 2. The power spectrum of Super-Kamiokande flux measurements over the frequency range 0 – 40 y$^{-1}$.

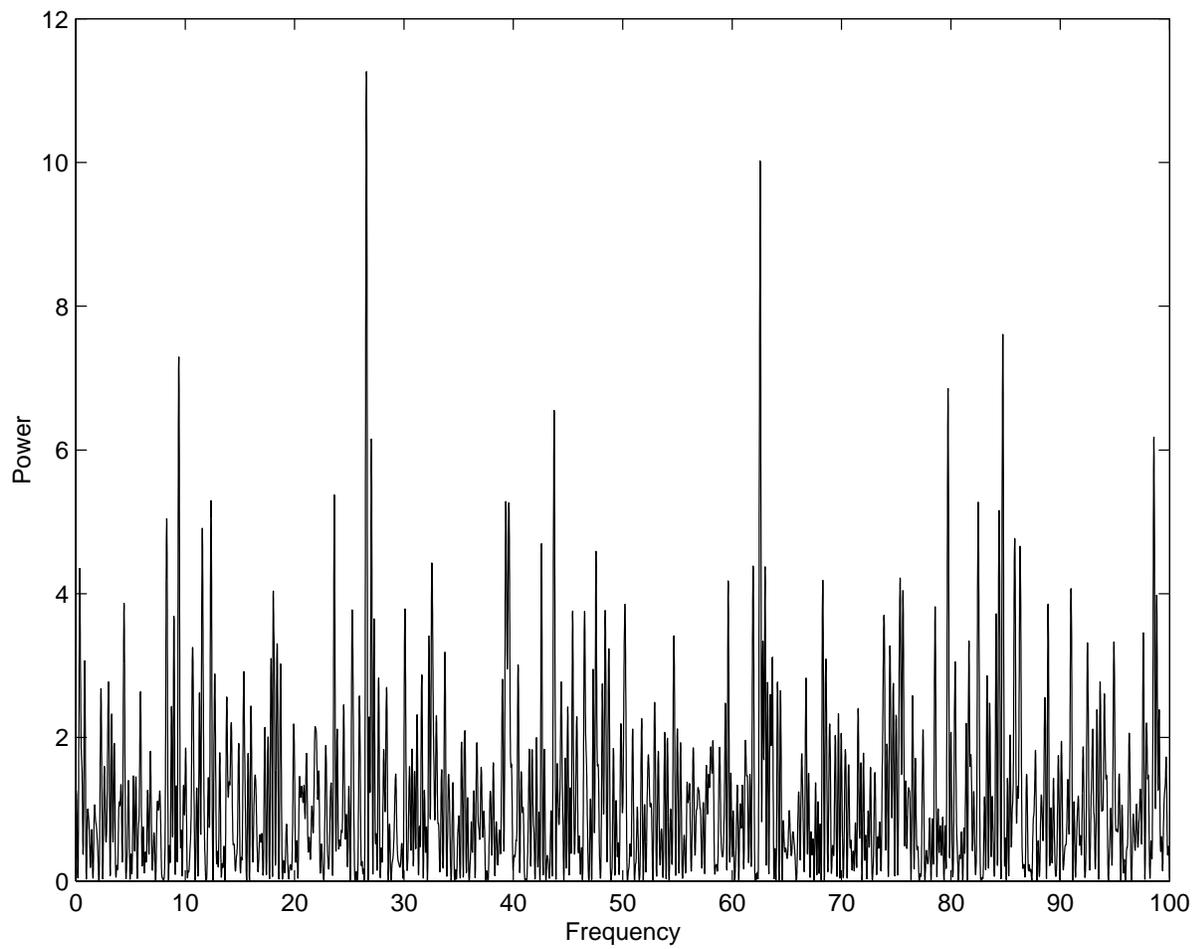

Figure 3. The power spectrum of Super-Kamiokande flux measurements over the frequency range 0 – 100 $y^{-1}$.

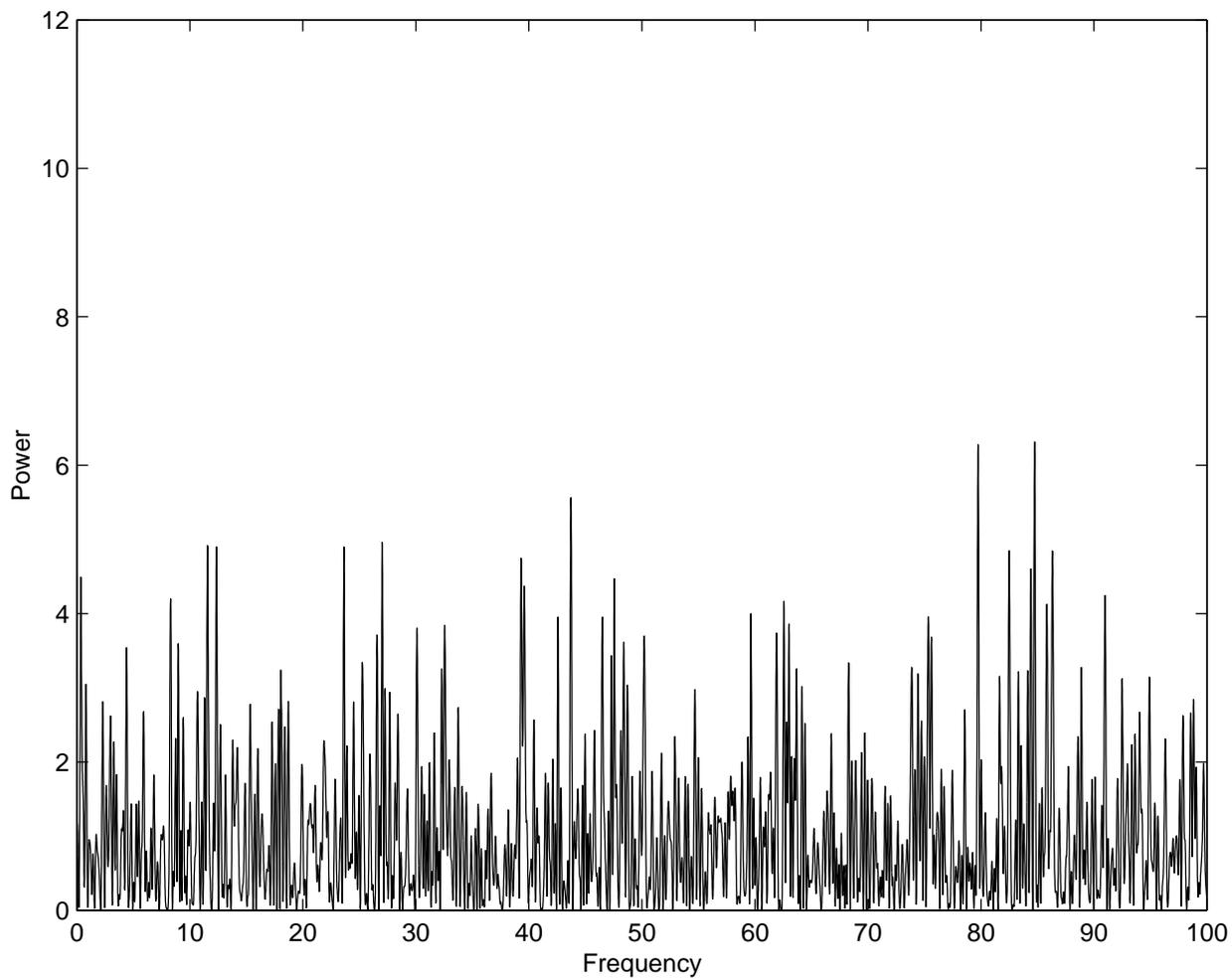

Figure 4. The power spectrum of Super-Kamiokande flux measurements over the frequency range 0 – 100 y$^{-1}$ after the oscillation at frequency 26.57 y$^{-1}$ has been removed.

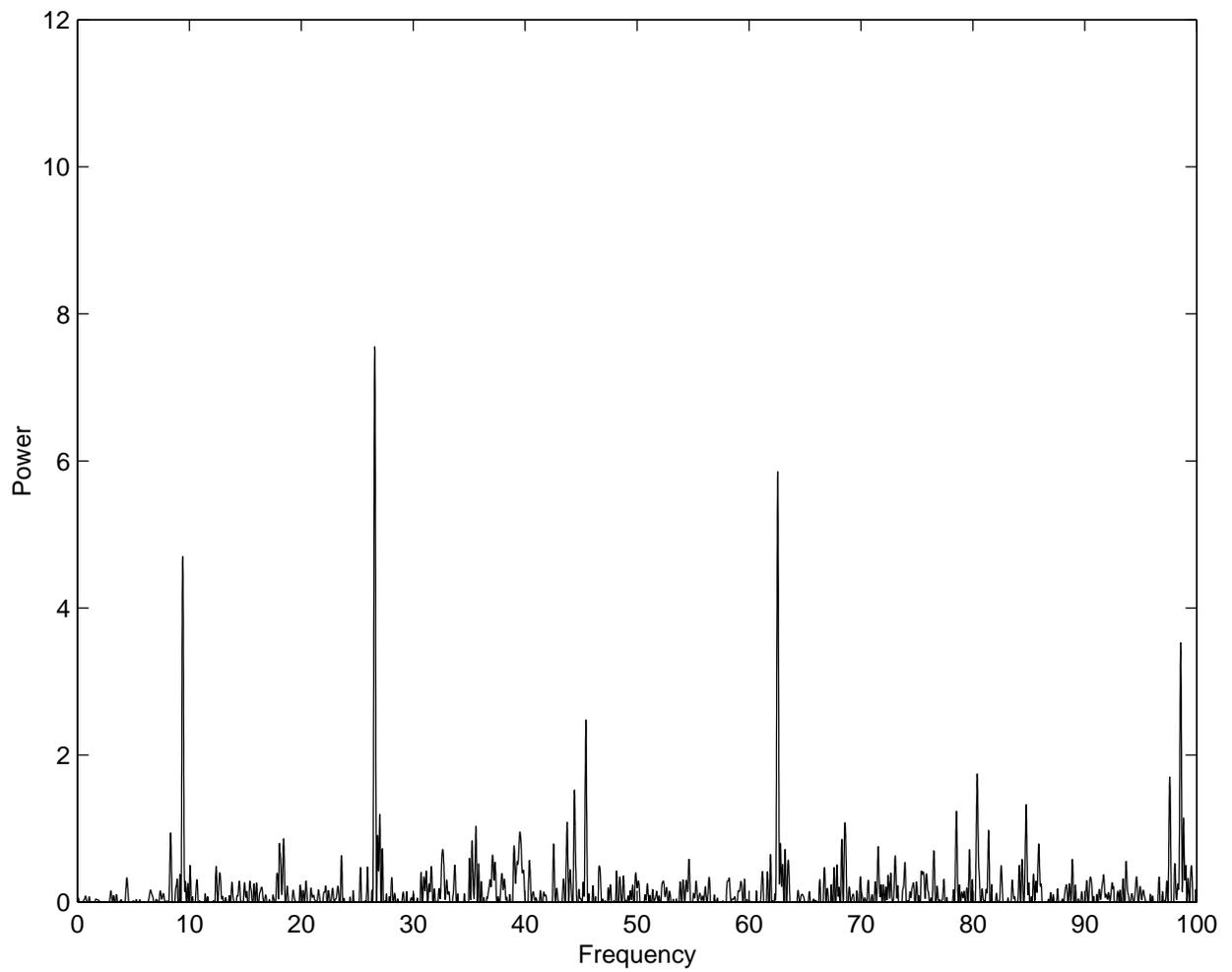

Figure 5. The difference between the power spectrum shown in Figure 3 and that shown in Figure 4.

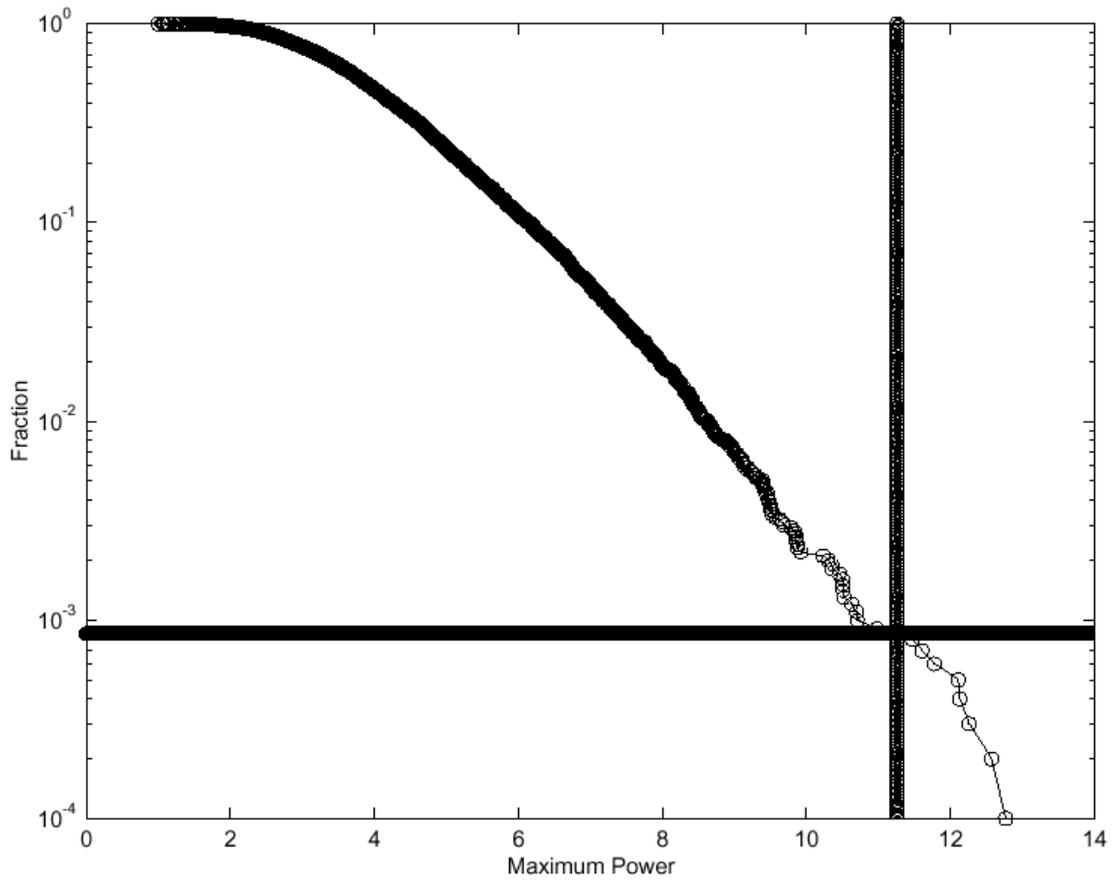

Figure 6. The ordinate is the fraction of 10,000 simulations that have maximum power, in the frequency range 25.36 to 27.66 $y^{-1}$, larger than the value indicated by the abscissa. The vertical line indicates the actual maximum power in that range. Less than 0.1% of the simulations have a maximum power larger than the actual maximum.

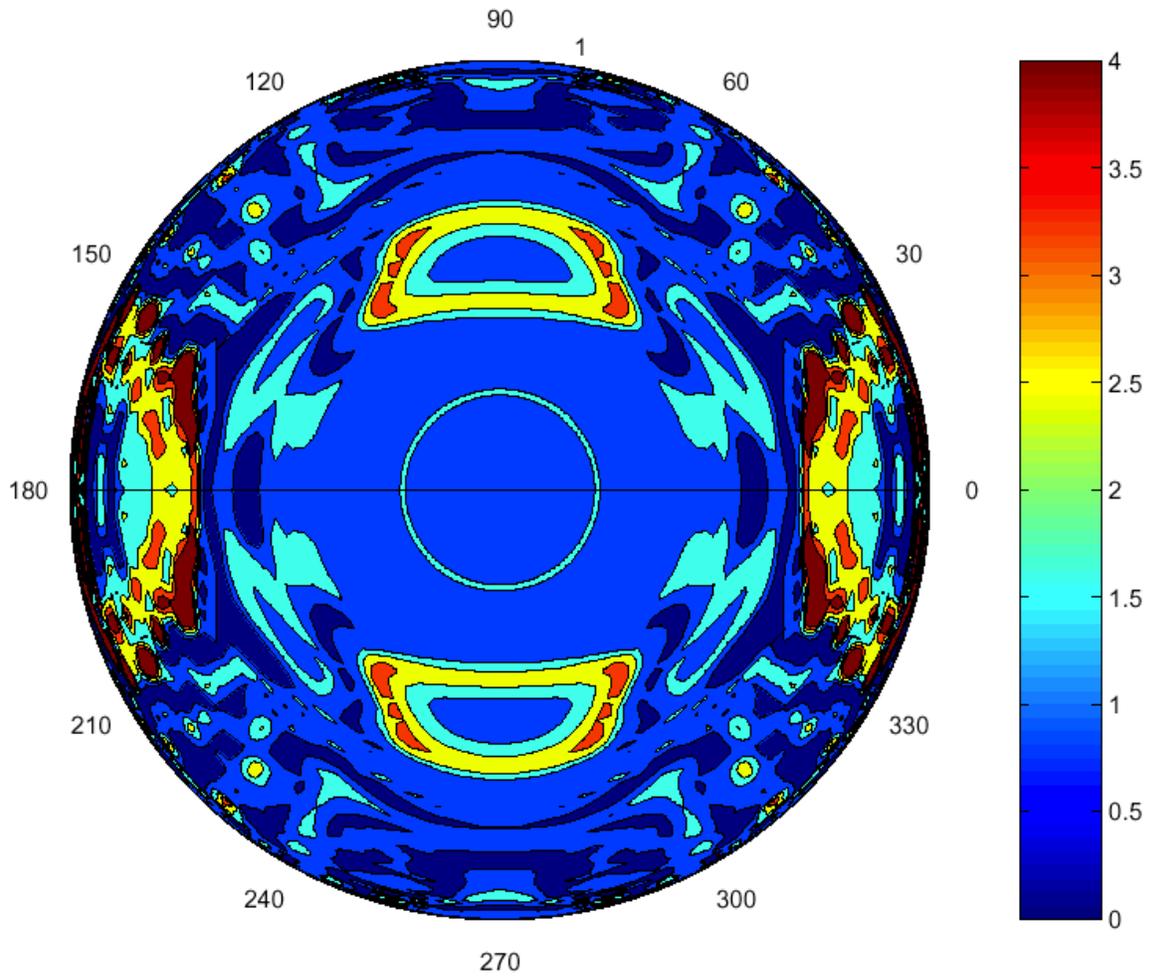

Figure 7. Map of the resonance statistic, defined by eq. (6.2), upon a meridional section of the solar interior. Red or yellow indicates a "resonance" between the neutrino flux and the harmonic of the local solar rotation.

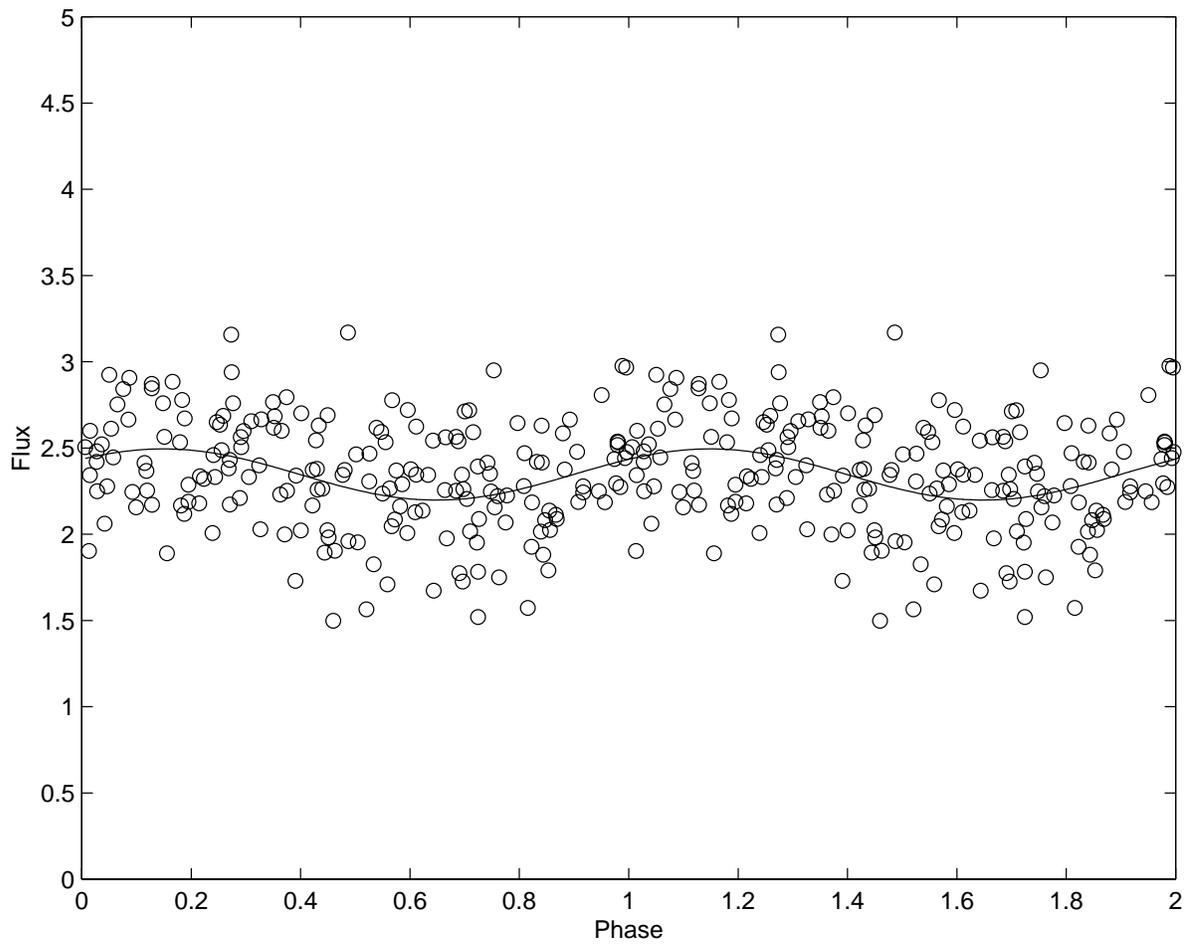

Figure 8. Plot of flux measurements (in units of $10^6$ cm$^{-2}$ s$^{-1}$) versus phase determined by the mean time of each bin and frequency 26.57 y$^{-1}$. Zero phase corresponds to the date 1970.00.

22